%% file: kleinjan.tex
\documentclass[12pt]{article}
\usepackage{graphicx}
\usepackage{units}
\usepackage{url}

\newcommand\pubnumber{CIPANP2015-Kleinjan}
\newcommand\pubdate{\today}

\def\napoli{Los Alamos National Laboratory\\
For the E906/E1039 Collaboration}

\def\Title#1{\begin{center} {\Large #1 } \end{center}}
\def\Author#1{\begin{center}{ \sc #1} \end{center}}
\def\Address#1{\begin{center}{ \it #1} \end{center}}

\newcommand\pubblock{\rightline{\begin{tabular}{l} \pubnumber\\
         \pubdate  \end{tabular}}}
\newenvironment{Abstract}{\begin{quotation}  }{\end{quotation}}
\newenvironment{Presented}{\begin{quotation} \begin{center} 
             PRESENTED AT\end{center}\bigskip 
      \begin{center}\begin{large}}{\end{large}\end{center} \end{quotation}}

\input econfmacros.tex

\begin{document}
\begin{titlepage}
\pubblock

\vfill
\Title{A Future Polarized Drell-Yan Experiment	
at Fermilab}
\vfill
\Author{ David Kleinjan}
\Address{\napoli}
\vfill
\begin{Abstract}
One of the great challenges of QCD is trying to understand the origin of the nucleon spin. Several decades of experimental measurements have shown that our current understanding is incomplete if only the quark and gluon spin contribution is considered. Over the last few years it has become increasingly clear that the contribution from the orbital angular momentum of the quarks and gluons has to be
included as well. For instance, the sea quark orbital contribution remains largely unexplored.  Measurements accessing the sea quark Sivers distribution will provide a probe of the sea quark orbital contribution. The upcoming E1039 experiment at Fermilab will access this distribution via the Drell-Yan process using a 120 GeV unpolarized proton beam directed on a polarized proton target. At E1039 kinematics the $u$-$\bar{u}$ annihilation process dominates the Drell-Yan cross section ($x_{Target}$ = 0.1 $\sim$ 0.35).  If the $\bar{u}$ quark carries zero net angular momentum, then the measured Drell-Yan single-spin asymmetry should be zero, and vice versa. This experiment is a continuation of the currently running SeaQuest experiment. 	
\end{Abstract}
\vfill
\begin{Presented}
Twelfth Conference on the Intersections of Particle and Nuclear Physics (CIPANP2015)\\
Vail, Colorado, May 19-24, 2015
\end{Presented}
\vfill
\end{titlepage}
\def\thefootnote{\fnsymbol{footnote}}
\setcounter{footnote}{0}

\section{Introduction\label{sec:intro}}
Our current understanding of the nucleon's spin is that it originates from the sum of the spin and orbital angular momentum of the partons confined in the nucleon
\begin{equation}
\displaystyle S_{n} = \nicefrac{1}{2} = \nicefrac{1}{2} \Delta \Sigma + \Delta G + L_{q,\bar{q},g}
\label{eq:spinsum}
\end{equation}

\noindent where $\Delta \Sigma$ is the net quark spin from all flavors, $\Delta G$ is the net gluon spin, and $L_{q,\bar{q},g}$ is the orbital angular momentum contribution from the respective quarks, anti-quarks, and gluons.  The net quark spin has been highly constrained by various experiments to be $\Delta \Sigma$\,$\approx$\,30$\%$ \cite{deltasigma}.  Recent results from RHIC indicate that the gluon spin contribution is possibly non-zero, $\Delta G$\,=\,20\,$\pm$\,12$\%$ \cite{rhicspin0}. To fully understand the nucleon spin, the orbital angular momentum contributions of the quarks and gluons must be considered as well.

Measurements accessing the orbital angular momentum of the valence quarks have been done by Semi-Inclusive Deep Inelastic Scattering experiments (SIDIS) at HERMES \cite{hermes0}, COMPASS \cite{compass0}, and JLAB \cite{jlab0}.  They have indicated that the angular momentum of up quarks is positive ($L_u$\,$>$\,0), while the down quark is negative ($L_d$\,$<$\,0), such that $L_{q}$\,=\,$L_{u}$\,+\,$L_{d}$\,$\approx$\,0 \cite{anselmino0}.  

\begin{figure}[h]
\begin{center}
\includegraphics[width=25pc]{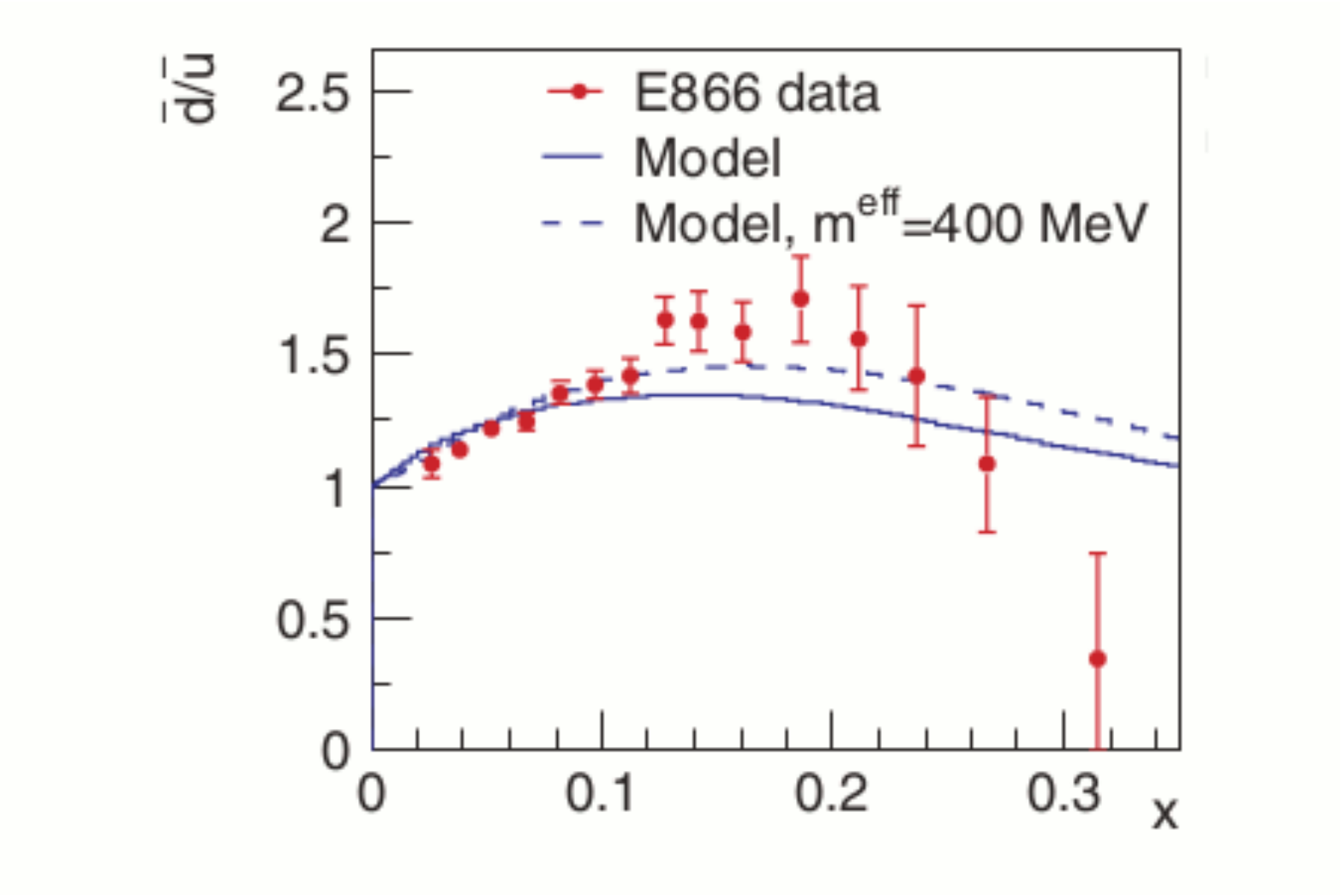}
\end{center}
\caption{\label{fig:e866} The $\bar{d}(x)/\bar{u}(x)$ ratio measured by the E866 collaboration \cite{e866}, compared to predictions from the pion cloud model \cite{pioncloud0}.}
\end{figure}

Measurements of the sea quark orbital angular momentum ($L_{\bar{q}}$) remain largely unexplored, as current SIDIS data have poor sensitivity to the sea quarks.
However, Lattice QCD calculations which predict the observed $L_{q}$\,$\approx$\,0, also predict up to 50$\%$ of the nucleon's spin to come from  $L_{\bar{q}}$ \cite{lattice0,lattice1}.  In addition to this theoretical prediction, indirect hints of a large non-zero $L_{\bar{q}}$ have been observed. Using proton induced Drell-Yan, the E866 experiment at Fermilab found an excess of $\bar{d}$ over $\bar{u}$ quarks, as seen in the $\bar{d}(x)/\bar{u}(x)$ ratio shown in Fig. \ref{fig:e866} \cite{e866}.  Several models can explain this $\bar{d}(x)/\bar{u}(x)$ ratio, among them the pion cloud model \cite{pioncloud0}.  The pion cloud  model describes the proton as a linear combination of a bare proton plus pion-baryon states.
\begin{equation}
\displaystyle |p \rangle = |p \rangle + | N^{0}{\pi}^{+} \rangle + | {\Delta}^{++}{\pi}^{-} \rangle + ...
\end{equation}

Since the proton is more likely to be in the pion-nucleon state than the pion-delta state, this leads to an excess of $\bar{d}$ versus $\bar{u}$ quarks.  Since the baryon has positive parity and the pion has negative parity, the pion cloud must be in an odd angular momentum state with respect to the baryon, \textit{i.e.} its substituent anti-quark will carry non-zero angular momentum, $L_{\bar{q}}$.  

\section{Accessing the Sea Quark Orbital Angular Momentum}

The nucleon's structure is studied and understood by measuring it's parton distribution functions (PDFs).  PDFs represent the probability that a given parton flavor carries a certain momentum fraction of the nucleon's momentum. By taking into account the spin of the proton with respect to the longitudinal momentum the proton, as well as spin-orbital properties of the partons, eight leading-twist transverse momentum dependent PDFs (TMDs) can be measured, which are summarized in \cite{jj_scheme, m_scheme}.


The different TMDs probe different properties of the proton.  Our focus will be upon the Sivers TMD function, as it directly correlates to the orbital angular momentum of the partons mentioned in Eq. \ref{eq:spinsum}.  It describes correlations between the transverse spin of the polarized nucleon and the transverse momentum of the parton with respect to the momentum of the nucleon \cite{sivers0,sivers1}.  

The Sivers function was originally formulated to explain the large transverse single spin asymmetries (SSAs) measured for inclusive hadron production in $p+p^{\uparrow}$ collisions \cite{e704}.  The quark Sivers function can be directly accessed from measured transverse SSAs in Polarized SIDIS $\left( e+p^{\uparrow} \rightarrow e+h+X \right)$ or Polarized Drell-Yan $\left( p+p^{\uparrow} \rightarrow {\gamma}^{*} \rightarrow \ell^{+} \ell^{-} \right)$.  The latter of these has never been carried out experimentally.  The same Sivers function is involved in both processes. However, transverse SSAs in SIDIS are the result of an attractive final state interaction whilst in Drell-Yan they are the result of a repulsive initial state interaction. This leads to a predicted sign change in the observed Sivers function
\begin{equation}
\displaystyle f^{\perp q}_{1T}|_{SIDIS} = -f^{\perp q}_{1T}|_{DY}
\end{equation}

Measured transverse SSAs in polarized SIDIS have been used to measure the quark Sivers function, which as already mentioned in Section \ref{sec:intro} has been used to show $L_{q}$\,=\,$L_{u}$\,+\,$L_{d}$\,$\approx$\,0 \cite{anselmino0}. However since valence and sea quarks cannot be isolated from each other in the SIDIS process, the sea quark Sivers function remains poorly constrained.

In measuring the transverse SSA from the polarized Drell-Yan process, the valence and sea quarks can be isolated from each other.  In addition, there are no fragmentation functions involved, so the transverse SSA of polarized Drell-Yan, or \textit{Sivers asymmetry}, is directly related to the Sivers function.  The E1039 experiment, described in the next section, will be the first to measure the sea quark Sivers function.

\section{The E1039 Experiment}

The E1039 experiment at Fermilab \cite{e1039} will measure the Sivers asymmetry of the Drell-Yan process from an unpolarized proton beam on a transversely polarized proton target via its decay into muons

\begin{equation}
\displaystyle A^{Sivers}_{N} \left( p_{beam} + p^{\uparrow}_{target} \rightarrow \gamma{*}\rightarrow \mu^{+}\mu^{-} \right) \propto \frac{1}{P_{target}}\frac{N^{DY}_L - N^{DY}_R}{N^{DY}_L + N^{DY}_R}
\label{eq:sivan}
\end{equation}

The dimuon pairs will be measured using the existing muon spectrometer used in the SeaQuest Drell-Yan experiment at Fermilab \cite{e906}. The 120 GeV Main Injector beam used for the SeaQuest experiment, with upgraded beam focusing, will also be used for E1039.  The new E1039 target will be set upstream from the spectrometer to optimize measurement of DY pair production from annihilation of a high momentum fraction valence quark from the beam, and a sea quark from the target with momentum fraction 0.1\,$<$\,$x_{\bar{q}}$\,$<$0.35.

In order to create the transversely polarized target, it must consist of some ensemble of polarizable protons in a high magnetic field ($B$) and cooled to low temperature ($T$). The \textit{thermal equilibrium} (TE) polarization of spin-$\nicefrac{1}{2}$ particles is described by 

\begin{equation}
\displaystyle P_{i} = tanh \left(\frac{g_i \mu_i B}{2 k_B T} \right)
\label{eq:tep}
\end{equation}

\noindent which is derived from Boltzmann's law and the Zeeman interaction of the particle's magnetic moment with the magnetic field using statistical mechanics.  Due to the small magnetic moment ($\mu_p$) of the proton, the TE polarization of protons at a low temperature of $T$\,=\,1~Kelvin and a high magnetic field of $B$\,=\,5~Tesla is only $P_{p}$\,$=$\,0.5$\%$.  However, electrons have a much higher magnetic moment ($\mu_e$\,$=$\,660$\mu_p$) and can be polarized up to $P_{e}$\,$=$\,99.8$\%$.  

\begin{figure}[h]
\begin{center}
\includegraphics[width=30pc]{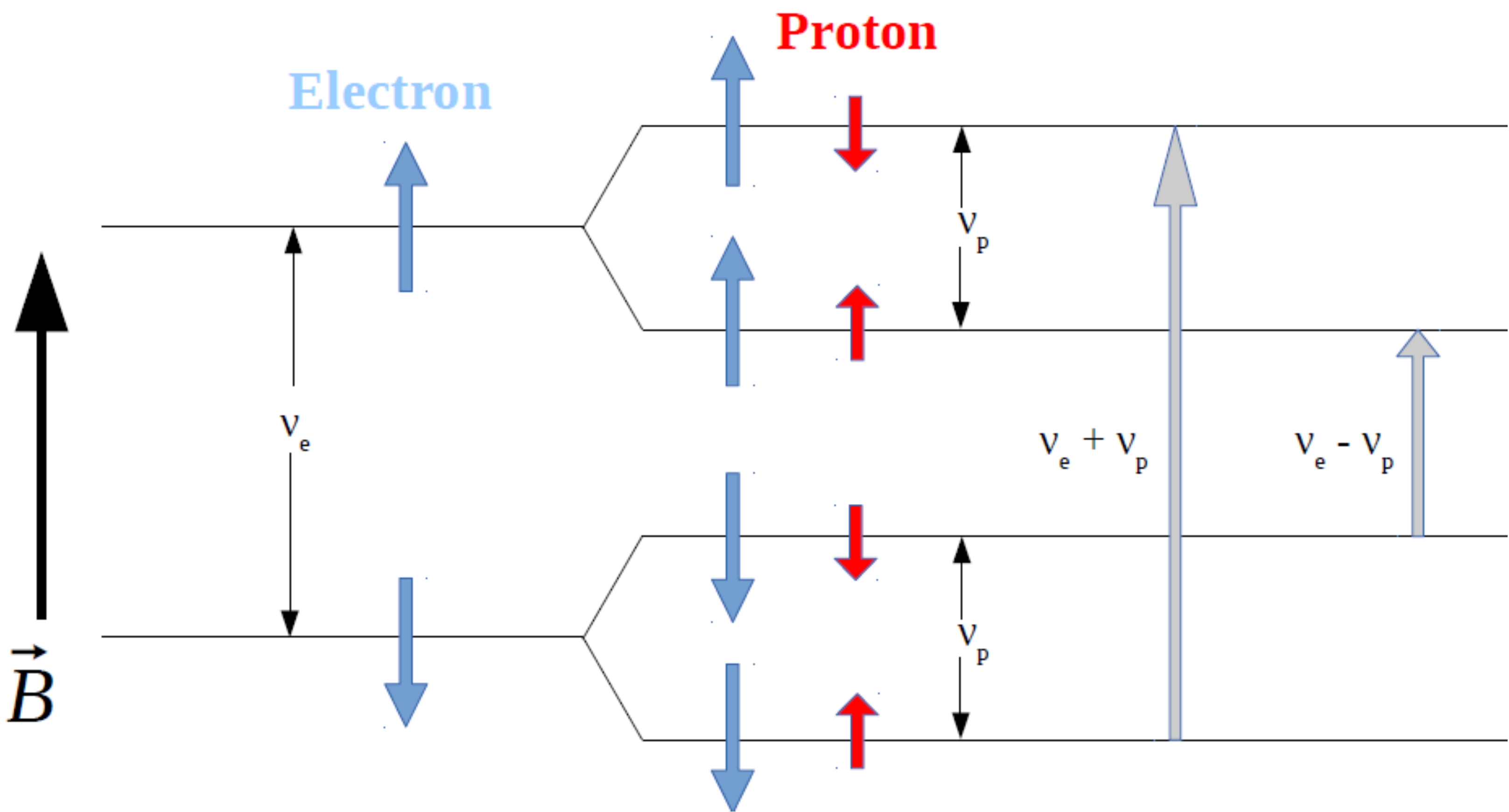}
\end{center}
\caption{\label{fig:dnp} Diagram indicating how polarization is passed from electron to nucleon via hyperfine splitting of the energy states \cite{Crabb}.}
\end{figure}

This high electron polarization can be transferred to the protons in a paramagnetic material utilizing \textit{Dynamic Nuclear Polarization} (DNP) \cite{dnp0}.  The dipole-dipole interaction between the proton and electron leads to hyperfine splitting, as shown in Fig. \ref{fig:dnp}. By providing an RF field equal to the sum (difference) of the electron and proton Larmor frequencies ($\nu_e \pm \nu_p$), the high polarization of the electron is transferred to the proton anti-aligned (aligned) with the magnetic field $B$ (see Fig. \ref{fig:dnp}). Due to the high relaxation time of the proton spin ($\tau_p$\,$>>$\,$\tau_e$), a polarization of the proton can be achieved comparable to the electron polarization.

For E1039, the polarized target will be made of irradiated ammonia $\left( NH_{3} \right)$. Irradiation of $NH_{3}$ is done to create paramagnetic centers, and DNP is used to polarize the protons in the hydrogen atoms of $NH_{3}$ \cite{nh30}.  When induced at a low temperature of $T$\,=\,1~Kelvin, and a high magnetic field of $B$\,=\,5~Tesla, polarization of the protons in $NH_{3}$ can reach up to $P$\,=\,92$\%$, which is measured using a \textit{Nuclear Magnetic Resonance} (NMR) technique \cite{NIM}.  

NMR works by applying an RF at the proton Larmor frequency ($\nu_p$) to a series RLC circuit, where the inductor is within the target material.  The polarization is measured by the absorption (emission) of RF in the circuit, which indicates that the polarization of proton's is aligned (anti-aligned) with the magnetic field.  The RF Voltage across the circuit increases (decreases) for absorption (emission).  The polarization can be measured as DNP is introduced to the target based on the known TE polarization

\begin{equation}
\displaystyle P_{DNP} = \frac{P_{TE}}{V_{TE}} V_{DNP}
\end{equation}

A liquid helium bath is used to cool superconducting coils which provide the 5-$T$ magnetic field to the $NH_{3}$ target. Liquid helium is also used in the refrigerator designed to cool the $NH_{3}$ target chamber. In addition, pumping on the liquid helium vapor in the target chamber lowers its vapor pressure, thus reducing the temperature of the $NH_{3}$ target to $\sim$1-K.  A full description of this type of magnet-refrigerator system to be used for E1039 can be found in \cite{Crabb}.

\section{Conclusion}
The measurement of the Sivers asymmetry by the E1039 experiment will provide a first look into the $\bar{u}$-quark Sivers function.
\begin{equation}
\displaystyle A^{Sivers}_{N} \left( p_{beam} + p^{\uparrow}_{target} \rightarrow \gamma{*} \rightarrow \mu^{+}\mu^{-} \right) \propto \frac{f^{u}_{1} \left(x_{beam} \right) \cdot f^{\perp ,\bar{u}}_{1T} \left(x_{target} \right)}{f^{u}_{1}\left( x_{beam} \right) \cdot f^{\bar{u}}_{1}\left( x_{target}\right)}
\end{equation}

\begin{figure}[h!]
\begin{center}
\includegraphics[width=24pc]{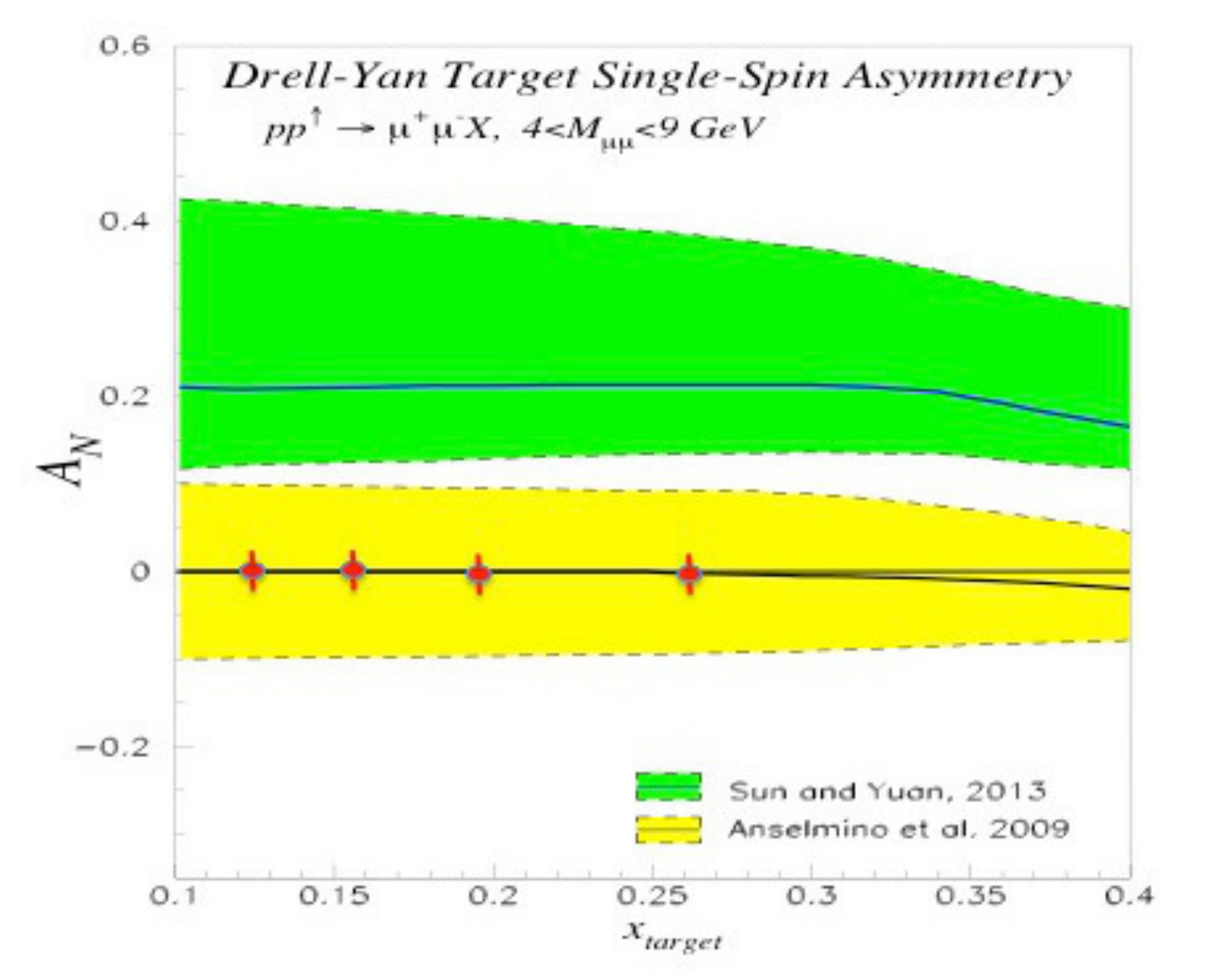}
\end{center}
\caption{\label{fig:proj} The projected error on the measurement of $A^{Sivers}_N$ vs. Bjorken-$x$ for one year of running based on annual POT\,=\,2.6\,$\times$\,$10^{18}$, compared to two predictions \cite{anselmino1,yuan0}.}
\end{figure}

Based on the projected beam luminosity and the current target, accelerator, and muon spectrometer efficiencies of SeaQuest, one year of running at E1039 will provide $\sim$2.6\,$\times$\,$10^{18}$ Protons on Target (POT).  Using this value and assuming an average polarization of $P$\,=\,80$\%$, Fig. \ref{fig:proj} gives the projected error on the Sivers asymmetry measurement.  The current plan is that E1039 will run for two years.

Combined with QCD theory, this experiment will provide the first measurement of the sign and magnitude of the sea-quark orbital angular momentum within the proton.  If $A^{Sivers}_N$\,$\neq$\,0, it will be the first experimental evidence to show that $L_{sea}$\,$\neq$\,0, a crucial piece of the nucleon spin puzzle.  Equally interesting is if $A^{Sivers}_N$ is consistent with zero. If $A^{Sivers}_N$\,=\,0, the observed $\bar{d}\left(x\right) /\bar{u}\left(x\right)$ flavor asymmetry seen by E866 \cite{e866}, and the origin of the nucleon spin, will remain a mystery.  Regardless of the findings, the measurement of $A^{Sivers}_N$ at E1039 is sure to provide new insights into nature of the nucleon.


\bibliographystyle{aipprocl} 
\bibliography{kleinjan}

\end{document}

%% file: econfmacros.tex
\def\beq{\begin{equation}}
\def\eeq#1{\label{#1}\end{equation}}
\def\eeqn{\end{equation}}

\def\beqa{\begin{eqnarray}}
\def\eeqa#1{\label{#1}\end{eqnarray}}
\def\eeqan{\end{eqnarray}}

\let\bar=\overbar

\def\Dslash{\not{\hbox{\kern-4pt $D$}}}
\def\dslash{\not{\hbox{\kern-2pt $\del$}}}

\def\msb{{\bar{\ssstyle M \kern -1pt S}}}

